\documentclass[legalpaper]{aa}

\usepackage{graphics}

\input{epsf.tex}

\begin{document}

\title{The velocity dispersion profile of globular clusters : a closer look}
\titlerunning{Velocity dispersion : a closer look}

\author{Fran{\c c}ois Roueff \inst{1}
\and Pierre Salati \inst{1,2}
\and Richard Taillet \inst{3}}

\institute{Laboratoire de Physique Th{\'e}orique ENSLAPP, BP110, F-74941
Annecy-le-Vieux Cedex, France.
\and Universit{\'e} de Savoie, BP1104 73011 Chamb{\'e}ry Cedex, France.
\and The Center for Particle Astrophysics, Le Conte Hall, University
of California at Berkeley, Berkeley, CA, 94720, USA.}

\thesaurus{10(10.07.02; 12.04.1)}

\offprints{Pierre Salati} \mail{Pierre Salati}

\date{Received / Accepted }

\maketitle

\begin{abstract}
Measurements of the surface brightness distribution and of the velocity
dispersion profile have been so far used to infer the inner dynamics of
globular clusters. We show that those observations do not trace back the
dark matter potentially concealed in these systems in the form of
low-mass compact objects. We have built Michie
models of globular clusters which contain both massive and low-mass stars.
An analytic expression for the stellar mass densities has been explicitely
derived in terms of the usual error function and Dawson's integral.
While the heavy population is kept fixed, the abundance of the light species
of our models is varied. When stellar velocities are anisotropic, both
the surface brightness and the velocity dispersion profiles of the cluster
become insensitive to the abundance of low-mass stars. This suggests that
the actual stellar mass function of many globular clusters is still to be
discovered.
\end{abstract}

\section{Introduction}

Among the different kinds of astrophysical star conglomerates, globular
clusters (GC) have a very peculiar status. First, as we go down the mass-scale
of observed structures, from super clusters through galaxies down to dwarf
spheroidals, they are the first objects whose dynamics can be understood
without resorting to the presence of dark matter. Second, and this may be
related to the first point, they are the only star clusters dense enough for a
statistical equilibrium to be reached in a Hubble timescale. The dark matter
content of globular clusters may be an important clue to the understanding
of the formation of our galaxy. However, the observational situation is not
very clear yet as regards stellar counts and the related direct detection of a
hidden population of light and faint stars. Since the claim by Richer and
Fahlman that the mass function steeply rises towards low masses in some
GCs
%\cite{Fahlman89,Richer91}
(Fahlman et al. 1989, Richer et al. 1991), the Hubble Space Telescope (HST)
has been used to measure the mass function in several globular clusters, and the
opposite conclusion was reached
%\cite{Demarchi95,Elson95,Paresce95}
(Demarchi and Paresce 1995, Elson et al. 1995, Paresce, De Marchi and Romaniello
1995).
Part of the discrepancy is due to the conversion scheme used to translate
the measured luminosity functions into mass functions. There is actually
no consensus on the  mass-to-luminosity relation of low-mass objects.
In a recent investigation,
%\cite{Santiago96}
Santiago et al. (1996) found a steadily
increasing mass function for \object{$\omega$ {\sl Cen}}, using HST
observations. 
This suggests that the dark matter content may depend on the cluster.
It seems therefore important to fully understand what we can also learn
from other types of measurement, such as in particular the brightness and
the velocity dispersion profiles.

There is a widespread belief that globular clusters cannot contain large
amounts of dark matter, because of the Virial theorem. The latter relates
the velocity dispersion $\sigma$ at the center of the cluster to the total
mass $M_\mathrm{t}$ and to the half-mass radius $r_\mathrm{h}$ of the
system. For a 
one-component globular cluster, that relation may be expressed as
%\cite{Spitzer87}
(Spitzer 1987) :
\begin{eqnarray}
\langle \sigma^2 \rangle \approx 0.4 \frac{G M_\mathrm{t}}{r_\mathrm{h}}
\label{viriel}
\end{eqnarray} 
However, if a dark component is also present, in the form of low-mass
objects for example, the virial theorem must be modified to
\begin{eqnarray}  
\sum_{i} M_\mathrm{t}(i) \langle \sigma^2 \rangle_{i} \; = \; E_p \;\; ,
\end{eqnarray}
where the index $i$ refers to the different stellar species, and where $E_{p}$
is the potential energy of the cluster. In previous papers
%\cite{Taillet95,Taillet96}
(Taillet et al. 1995 and 1996)
focusing on isotropic globular clusters, we showed that in the presence of dark
matter, relation (\ref{viriel}) still holds for the visible component, and that
most of the visible properties of the cluster are approximately the same as for
a one-component system. In particular, the brightness profile is nearly unchanged
even if dark matter is abundant. This is due to the phenomenon of mass
segregation. Because of gravitational interactions, stars tend to share their kinetic
energy. Thus, the light objects gain velocity from the heavier stars, which sink to
the center of the cluster, while the light objects have a much more diffuse
distribution. The heavier stars then form a self-gravitating system in the
low-density background of the light objects, and their properties are not
significantly altered. It was concluded that even a dark-matter dominated
cluster looks very much the same as a dark-matter free system, and that the point
of the dark matter content could not be made by the mere observation of the
brightness profile.

It was also noted that the presence of dark matter would be betrayed
by the velocity dispersion profile of bright stars which flattens when dark matter
is present. However, those profiles need to be accurately measured, which is not
yet the case. The results of our analysis were obtained for King models where
the stellar distribution is isotropic in velocity space. Their validity must be
reassessed when the velocity distribution is no longer isotropic. Anisotropy is
likely to be present at some level because stars mainly interact in the cluster core.
At least part of those orbiting at the outskirts of the system gained the required
energy by interacting at the center. Then, orbits should be predominantly radial
at large radii. In that respect, note that \object{$\omega$ {\sl Cen}}
for which the HST 
observations provide a steadily rising mass function may also have an
anisotropic velocity distribution
%\cite{Meylan87}
(Meylan 1987), even though an isotropic distribution is not excluded
(Merrit et al. 1996). Such a cluster may be modeled by a Michie
distribution in phase space, where an angular-momentum exponential cut-off
is applied to a King distribution (see section~2).

It is well known that when anistropy is present, the velocity dispersion profile
is less flat. Thus we suspect that the flattening of the velocity dispersion profile
that occurs in the presence of dark matter will be partially cancelled by the effect
of anisotropy. A given velocity dispersion profile could therefore be interpreted
either with a one-component isotropic GC or with a two-component anisotropic
GC. However, anisotropy may also affect a priori the brightness profile. The main
goal of this paper is therefore to make these statements more quantitative. We
will investigate how anisotropy affects the detectability of dark matter in globular
clusters. In section~2, we present a new analytic form for the equation that
relates the gravitational potential to the stellar mass densities. The Michie
distribution is expressed with the usual error function and Dawson's integral.
Section~3 is devoted to our set of two-component Michie models whose heavy
component is kept fixed while the abundance in low-mass stars is varied. Various
velocity anisotropies are explored. Finally, the surface brightness and the velocity
dispersion profiles of these models are discussed in section~4 and conclusions
are drawn.

\section{The Michie distribution functions}

\begin{figure}
\resizebox{\hsize}{!}{\includegraphics{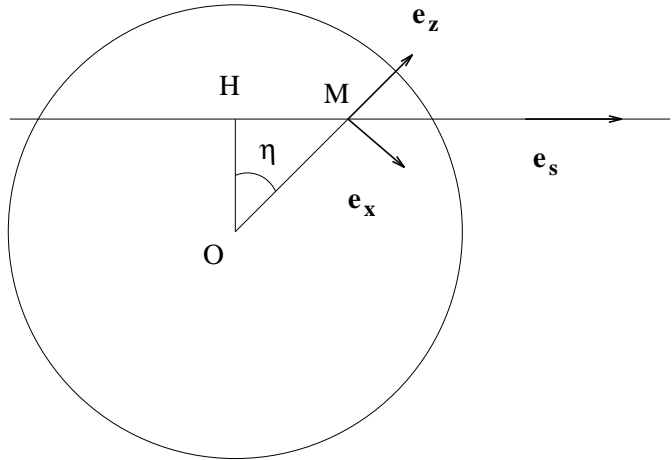}}
\caption{The unit vector $\vec{e}_\mathrm{s}$ is aligned along the
line of sight HM 
that crosses the globular cluster. At the pericluster H, the distance to
the center O is minimal. At any given point M on the line of sight,
with position labelled by the angle $\eta$, a convenient frame may
be defined as presented here.}
\label{figure1}
\end{figure}

A Michie model of a globular cluster is spherically symmetric as regards its
distribution of stars in space. However, its velocity ellipsoid tends to be
elongated towards the center of the cluster for radii larger than some critical
value $r_a$. The phase space distribution of a stellar species with mass
$m$ and whose one-dimensional velocity dispersion is $\sigma$, may be
expressed as
\begin{eqnarray}
f(r,\vec{v}) \; = \;
%\frac{n_{1}}{\left( 2 \pi \sigma^2 \right)^{3/2}} \;
k \;
e^{\displaystyle - L^2 / 2 r_a^2 \sigma^2} \;
\left\{e^{\displaystyle {\cal E} / \sigma^2} \, - \, 1 \right\}
\;\; ,
\label{MICHIE_1}
\end{eqnarray}
where $\vec{L} =  \vec{OM} \wedge \vec{v}$ is the orbital momentum
of the stars located at point M and whose velocities are $\vec{v}$.
That orbital momentum is defined with respect to the cluster center O.
The quantity $\cal{E}$ is related to the stellar energy per unit mass
$E = \Phi(r) + v^{2}/2$ through
\begin{eqnarray}
{\cal E} \; = \; \Phi_\mathrm{t} - E \;\; ,
\end{eqnarray}
where $\Phi_\mathrm{t} = \Phi(r_\mathrm{t})$ is the gravitational potential at the tidal
boundary $r_\mathrm{t}$ of the cluster. Whenever $\cal{E}$ is negative, the
distribution function $f$ vanishes. Inside the radius of anisotropy $r_{a}$,
the distribution of stellar velocities is spherical. Beyond $r_{a}$, it
straightens in the direction of the cluster center. At remote distances,
trajectories are almost radial. This is a major difference with King models
for which velocities are spherically distributed. The mass density is now
given by a double integral. The orbital momentum
$L = r \, v \, \sin \theta$ depends actually on both the magnitude $v$ of
the velocity and on the angle $\theta$ between the latter and the radial direction
$\vec{OM}$. Defining the variables $x = v / \sqrt{2} \sigma$ and $y = \cos \theta$
yields a mass density
\begin{eqnarray}
\rho(r) \; = \; 4 \pi k \; m\, \left( 2 \sigma^2 \right)^{3/2} \,
{\cal H} (u,\alpha)
\end{eqnarray}
proportional to the double integral
\begin{eqnarray}
{\cal H} (u,\alpha) \; = \;
{\displaystyle \int_{0}^{1}} dy \,
{\displaystyle \int_{0}^{\displaystyle \sqrt{u}}} dx \, x^{2} \,
h(x,y) \;\; ,
\label{MICHIE_2}
\end{eqnarray}
where the function $h(x,y)$ is defined as
\begin{eqnarray}
h(x,y) \; = \;
e^{\displaystyle - \alpha^{2} x^{2} \left( 1 - y^{2} \right)} \,
\left\{ e^{\displaystyle \left( u - x^{2} \right)} \, - \, 1 \right\} \;\; .
\end{eqnarray}
The parameter $u = \left( \chi - \psi \right)$ is the difference between the
reduced potential $\chi = \Phi_\mathrm{t} / \sigma^2$ at the tidal
boundary $r_\mathrm{t}$ of the system
and its counterpart $\psi = \Phi(r) / \sigma^2$ at distance $r$ from the cluster center.
The ratio $r / r_{a}$ is denoted by $\alpha$. The scaling factor between the
mass density $\rho(r)$ and the integral (\ref{MICHIE_2}) may also be inferred
from the relation
\begin{eqnarray}
{\displaystyle \frac{\rho(r)}{\rho(0)}} \; = \;
{\displaystyle \frac{{\cal H} (u,\alpha)}{{\cal H} (\chi,0)}} \;\; ,
\end{eqnarray}
where $\rho(0)$ denotes the stellar mass density at the cluster center.

So far, expression (\ref{MICHIE_2}) was estimated numerically
through a direct integration. We have derived here an analytic
development of ${\cal H} (u,\alpha)$ in terms of the functions
\begin{eqnarray}
d(x)  =  e^{\displaystyle - x^{2}} \,
{\displaystyle \int_{0}^{x}} \, e^{\displaystyle t^{2}} \, dt
\end{eqnarray}
and
\begin{eqnarray}
e(x)  =  e^{\displaystyle x^{2}} \,
{\displaystyle \int_{0}^{x}} \, e^{\displaystyle - t^{2}} \, dt \;\; .
\end{eqnarray}
The first integral is called the Dawson's function and may be
easily computed thanks to a convenient approximation due to Rybicki
%\cite{Rybicki89,NumRec}
(Rybicki 1989 and Numerical Recipes). The function $e(x)$ is related
to the error function
\begin{eqnarray}
e(x) \; = \; \frac{\sqrt{\pi}}{2} \;
e^{\displaystyle x^{2}} \; {\rm erf}(x) \;\; .
\end{eqnarray}
The distribution function $f(r,\vec{v})$ can be averaged over the
angle $\theta$ to yield a King distribution up to a correction factor
\begin{eqnarray}
f(r,v) \; = \; k \;
\left\{e^{\displaystyle {\cal E} / \sigma^2} \, - \, 1 \right\} \;
\left\{ \frac{d(\alpha x)}{\alpha x} \right\} \;\; .
\label{MICHIE_KING}
\end{eqnarray}
The parameter $\alpha x$ stands for the ratio $r v / \sqrt{2} r_{a} \sigma$.
Inside the radius of anisotropy, $\alpha x$ is small compared to unity
and the correction factor $d(\alpha x) / \alpha x$ of relation
(\ref{MICHIE_KING}) reduces to 1. A King distribution is therefore
recovered when the radius $r$ is smaller than $r_{a}$.
%a situation for which velocities are spherically distributed.
After an integration on the magnitude
$v$ of the velocity, relation (\ref{MICHIE_2}) may be expressed as
\begin{eqnarray}
{\cal H} (u,\alpha)  = 
\frac{1}{2 (1 + \alpha^{2})} 
\left[ e(\sqrt{u}) - \sqrt{u}  + 
\frac{d(\alpha \sqrt{u}) - \alpha \sqrt{u}}{\alpha^{3}}
\right]
\label{H_u_a}
\end{eqnarray}
Several approximations to that expression are given in the Appendix
depending on whether $\sqrt{u}$ and $\alpha \sqrt{u}$ are large or not
with respect to unity.

The dispersion velocity profile of globular clusters may be defined
in various ways. What astronomers actually measure is the velocity
of some stellar species along the line of sight. Suppose that a specific
pixel contains $N$ stars with same spectral type. The radial velocity
of the i$^{\rm th}$ object of the sample is denoted by $v_{i}$. A variety
of statistical averages can be performed, each yielding a different value
for the dispersion. In this article, the dispersion velocity
$\bar{v}_\mathrm{rad}$ along the line of sight will be defined as
\begin{eqnarray}
\bar{v}_\mathrm{rad} \; = \; \left\{
{\displaystyle \sum_{i = 1}^{N}} \, \frac{v_{i}^{2}}{N} \right\}^{1/2} \;\; .
\label{VR_OBS}
\end{eqnarray}
In our two component models, we will be interested in the velocity dispersion
profile of the heavy and bright stellar population for which spectral
measurements can be performed.

Let us consider now a line of sight with direction set by the unit vector
$\vec{e}_{\rm s}$ as shown in Fig.~\ref{figure1}. At the
pericluster H, the distance to the center
O is minimal. For any given point M along that line of sight, a convenient
frame may be defined with its unit vector $\vec{e}_{\rm z}$
pointing outwards in the 
same direction as the vector $\vec{OM}$. The unit vector
$\vec{e}_{\rm x}$ is 
perpendicular to $\vec{e}_{\rm z}$ and is in the plane defined by the
three points O, H and M. Finally, the last basis vector
$\vec{e}_{\rm y}$ is defined as the product 
$\vec{e}_{\rm z} \wedge \vec{e}_{\rm x}$. The position of point M may
be traced by the angle 
$\eta$ between the vectors $\vec{OH}$ and $\vec{OM}$. The projection
$v_{s}$ of the velocity along the line of sight is then given by
\begin{eqnarray}
\vec{e}_{\rm s} \cdot \vec{v} \; = \;
v_{x} \cos \eta \, + \, v_{z} \sin \eta \;\; .
\end{eqnarray}
Its square $v_s^2$ may be averaged locally to yield
\begin{eqnarray}
\left\langle v_s^2 \right\rangle \; = \;
\left\langle v_x^2 \right\rangle \cos^{2} \eta \, + \,
\left\langle v_z^2 \right\rangle \sin^{2} \eta \;\; .
\end{eqnarray}
Because the velocity distribution is axisymmetric around the direction
connecting point M to the cluster center O, the contribution of the
product $v_{x} v_{z}$ to the local average of the line of sight velocity
$v_{s}^{2}$ vanishes. The latter may be expressed in terms of the functions
${\cal I} (u,\alpha)$ and ${\cal J} (u,\alpha)$
\begin{eqnarray}
\rho (M) \,
\left\langle \left( \frac{v_s}{\sigma} \right)^2 \right\rangle
=  4 \pi k \; m\, \left( 2 \sigma^2 \right)^{3/2} \;
\left\{ {\cal I} + {\cal J} \, \sin^2 \eta \right\} 
\end{eqnarray}
where $\rho (M)$ denotes the mass density of the stellar species with
mass $m$ under consideration at the point $M$. The functions ${\cal I}$ and ${\cal J}$ are
respectively defined by the integrals
\begin{eqnarray}
{\cal I} (u,\alpha) \; = \;
{\displaystyle \int_{0}^{1}} dy \,
{\displaystyle \int_{0}^{\displaystyle \sqrt{u}}} dx \, x^{4} \,
\left( 1 - y^{2} \right) \, h(x,y) \;\; ,
\label{ISO_I}
\end{eqnarray}
and
\begin{eqnarray}
{\cal J} (u,\alpha) \; = \;
{\displaystyle \int_{0}^{1}} dy \,
{\displaystyle \int_{0}^{\displaystyle \sqrt{u}}} dx \, x^{4} \,
\left( 3 y^{2} - 1 \right) \, h(x,y) \;\; .
\label{ANI_J}
\end{eqnarray}
The functions ${\cal H}$ and ${\cal I}$ are related by
\begin{eqnarray}
{\cal I}(u,\alpha) \; = \; - \,
\frac{\partial {\cal H}}{\partial \alpha^{2}} \;\; ,
\end{eqnarray}
so that ${\cal I}$ may be expressed as
\begin{eqnarray}
\left( 1 + \alpha^{2} \right) \, {\cal I}(u,\alpha) &=&
{\cal H} \, + \, \frac{3}{4 \alpha^{5}}
\left\{ d \left( \alpha \sqrt{u} \right) - \alpha \sqrt{u} \right\}
\nonumber \\
&+& \frac{u}{2 \alpha^{3}} d \left( \alpha \sqrt{u} \right) \;\; .
\end{eqnarray}
The calculation of the function ${\cal J}$ is slightly more involved
and yields
\begin{eqnarray}
\left( 1 + \alpha^{2} \right)^{2} \, {\cal J}(u,\alpha) & = &
\frac{\alpha{2}}{2}
\left\{ e \left( \sqrt{u} \right) - \sqrt{u} \right\} \nonumber \\ 
&+& \left( \alpha^{2} + u + u \alpha^{2} \right) \frac{u^{3/2}}{3}
\nonumber \\
& - &
{\cal K}(u,\alpha)
\left\{ d \left( \alpha \sqrt{u} \right) - \alpha \sqrt{u} \right\}
\;\; ,
\end{eqnarray}
where
\begin{eqnarray}
{\cal K}(u,\alpha) \; = \;
\frac{2 \alpha^{2} \, + \,
\left( 1 + \alpha^{2} \right) \left( 5 + 2 u \alpha^{2} \right)}
{4 \alpha^{5}} \;\; .
\end{eqnarray}
Notice that $v_{s}^{2}$ needs eventually to be averaged along the
entire line of sight. The end result is therefore the radial average
\begin{eqnarray}
\bar{v}_\mathrm{rad}^2 \; = \; {\displaystyle
\frac{\displaystyle \int \, \rho (M) \,
\left\langle v_{s}^{2} \right\rangle \, ds}
{\displaystyle \int \, \rho (M) \, ds}
} \;\; ,
\label{velocity_dispersion}
\end{eqnarray}
to be compared with the observational definition (\ref{VR_OBS}).

\section{The models}

To test the sensitivity of observable quantities to the amount of dark
matter, we built two-component Michie models of globular
clusters, with both a heavy and a light stellar population. One of the
components encapsulates the visible, solar mass stars while the other
component is dark matter in the form of low-mass objects with mass
$m_{2} = 0.1\, \mbox{M}_\odot$. The indexes 1 and 2 respectively refer
to the bright 
and dark populations. We built 25 globular clusters, with anisotropy radii
$z_{a}$ = 1000, 100, 50, 20 and 10, and dark matter content
$M_{2}/M_{1}$ = 0, 1, 3, 6 and 10. The parameter $z_{a}$ stands for the ratio
of the anisotropy radius $r_{a}$ to the typical scale length $a$ yet to be
defined. Models with $z_{a}$ = 1000 may be considered as King models.
In that case, the anisotropy radius is so large that it contains most of the
cluster, hence a spherical velocity distribution almost everywhere.
A two-component Michie model is completely specified by six parameters,
namely the anisotropy radius $r_{a}$,
the velocity dispersions $\sigma_1$ and $\sigma_2$ of the two species,
the normalization factors $k_1$ and $k_2$
(or alternatively the central densities $\rho_{c1}$ and $\rho_{c2}$), and
finally the depth of the gravitational potential well
$\chi = \Phi_\mathrm{t} / \sigma_1^2$.
These parameters are entirely determined by the following conditions~:
\begin{itemize}
\item{(i)} The velocity dispersion for the luminous stars is measured.
We will adopt the typical value $\sigma_1 = 7 \, \mbox{km/s}$.
\item{(ii)} The two species are in thermal equilibrium, imposing the
relation $m_1 \sigma_1^2 \, = \, m_2 \sigma_2^2 $ which sets $\sigma_2$
equal to $\sqrt{10} \times \sigma_1$.
\item{(iii)} The visible central density is set to
$\rho_{c1} = 8000 \, \mbox{M}_\odot / \mbox{pc}^3$. This leads to models having
almost the same central brightness which, as a matter of fact, is a well
measured quantity.
\item{(iv)} The anisotropy radius has been varied as explained above.
The smaller $r_{a}$, the stronger the anisotropy.
\item{(v)} The total visible luminosity is also a well determined quantity.
Here, it has been set equal to a typical value $L_{1} = 3\times 10^{5}
\,\mbox{L}_\odot$. Throughout 
our set of models, $\rho_{c2}$ and $\chi$ have been chosen to yield that total
luminosity, as well as the above mentioned ratios $M_{2}/M_{1}$.
\end{itemize}
Remember
that $M_{1}$ and $M_{2}$ respectively stand for the total mass in heavy and in
low mass stars. Visible stars are assumed to have a standard solar $L/M$ ratio.
Low mass objects are too faint to shine and their luminosity is negligible.

\noindent The structure of the cluster is determined by the Poisson equation
where the local mass density is given by
\begin{eqnarray}
\rho (r) \; = \; \rho_{c1} \, \frac{{\cal H} (u , \alpha)}{{\cal H} (\chi , 0)}
\; + \; \rho_{c2} \,
\frac{{\cal H} (m_{2} u / m_{1} , \alpha)}{{\cal H} (m_{2} \chi / m_{1} , 0)} \;\; .
\end{eqnarray}
Introducing the dimensionless radius $z = r / a$ where the typical scale length
$a$ is defined as
\begin{eqnarray}
a \; = \; \frac{\sigma_1}{\sqrt{4\pi G \rho_{c1}}} \;\; ,
\end{eqnarray}
leads to the differential equation
\begin{eqnarray}
\frac{1}{z^{2}} \frac{d}{dz} \left( z^{2} \frac{du}{dz} \right) \; = \;
- \, \frac{{\cal H} (u , \alpha)}{{\cal H} (\chi , 0)} \, - \,
\left( \frac{\rho_{c2}}{\rho_{c1}} \right)
\frac{{\cal H} (m_2 u / m_1 , \alpha)}{{\cal H} (m_2 \chi /  m_1 , 0)}
\nonumber
\end{eqnarray}
Once that equation is solved and the structure of the cluster is determined,
the velocity dispersion and the surface brightness profiles are computed.
The former is evaluated with the help of relation (\ref{velocity_dispersion}).
The latter is just the surface mass density of heavy stars
\begin{eqnarray}
\Sigma_1 \; = \; {\displaystyle \int} \, \rho_1(m) ds \;\; ,
\end{eqnarray}
up to the factor $(L_{\odot} / M_{\odot}) / 4 \pi$.

\section{Discussion and conclusions}

\begin{figure}
\resizebox{\hsize}{!}{\includegraphics{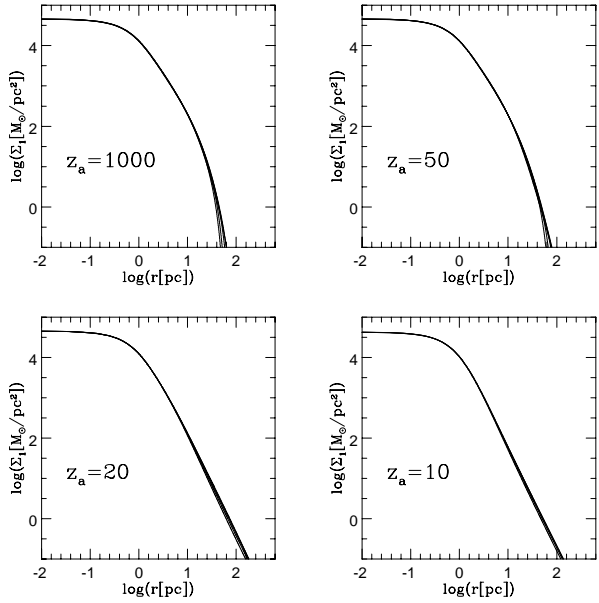}}
\caption{In each of the four plots, the surface mass density $\Sigma_1$
of the heavy population of our models is presented as a function of the
radius $r$. The mass ratio $M_{2} / M_{1}$ is set equal to the five different values
0, 1, 3, 6 and 10. The radius of anisotropy $z_{a}$ decreases from 1000, where a
King model is recovered with a spherical distribution of velocity, down to 10
for which the anisotropy is strong and the stellar orbits almost radial. In each plot,
the curves are hard to disentangle. The surface brightness profile is insensitive
to the dark matter content $M_{2}$ of the cluster.}
\label{figure2}
\end{figure}

\begin{figure}
\resizebox{\hsize}{!}{\includegraphics{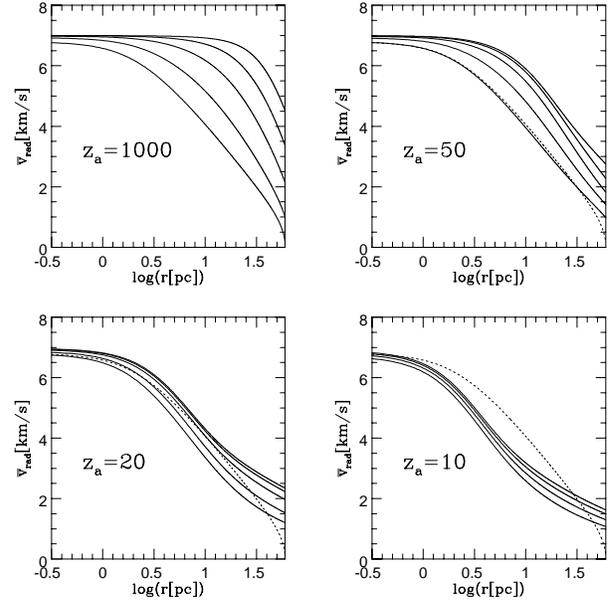}}
\caption{Same as in figure~2 with the velocity dispersion profiles.
For $z_{a}$ = 1000, velocities are almost spherically distributed and the
curves are different when the dark matter abundance $M_{2} / M_{1}$ varies.
However, the velocity dispersion along the line of sight becomes insensitive
to the dark matter content of the system when the anisotropy of the
velocity distribution increases. The dashed curve that appears in three of the
plots is the velocity dispersion profile for a King model with no low mass stars.
That case is degenerate with the entire set of Michie clusters with $z_{a} = 20$.
The same velocity dispersion profile may therefore be interpreted
either with a one-component isotropic cluster or, alternatively, with an
anisotropic system containing dark matter.
}
\label{figure3}
\end{figure}

Profiles of the surface mass density $\Sigma_1$ and of the
velocity dispersion $\bar{v}_\mathrm{rad}$ for the heavy component of
our models 
are presented in Fig.~\ref{figure2} and \ref{figure3} respectively. Low mass
stars do not shine in the V band and therefore do not contribute to
the surface luminosity of the cluster.
Figure~2 shows that whatever the anisotropy radius $z_{a}$, the
brightness profiles are almost not affected by the dark matter content
of the cluster. At constant $z_{a}$, the various curves are superimposed
while the light to heavy component mass ratio $M_{2} / M_{1}$ increases
from 0 to 10. This result was already obtained in the case of King models
%\cite{Taillet95,Taillet96}
(Taillet et al. 1995 and 1996) and its validity is now extended to
Michie clusters. For $z_{a}$ = 10 and 20, the surface luminosity
drops like $r^{- 5/2}$ as expected for a strongly anisotropic system.
In Fig.~\ref{figure3}, the velocity dispersion profiles are presented.
For $z_{a}$ = 1000, velocities are almost spherically distributed and the
curves are different when the dark matter abundance $M_{2} / M_{1}$ varies.
However, the velocity dispersion along the line of sight becomes insensitive
to the dark matter content of the cluster when the anisotropy of the
velocity distribution increases. For $z_{a}$ smaller than 20, the various
profiles start to be fairly similar and become hard to disentangle
observationally. Beyond the radius of anisotropy $r_{a}$, the stellar orbital
velocity $v_{\theta} \, = \, v \, \sin \theta$ is dominated on average
by the Michie exponential cut-off on the orbital momentum. The orbital
velocity is in that case of order
$v_{\theta} \sim \sigma \, r_a / r$.
It is no longer sensitive to the gravitational potential. The velocity dispersion
$\bar{v}_\mathrm{rad}$ measured along the line of sight, to which the
orbital velocity 
contributes most, is therefore blind to the mass content of the cluster.
As $z_{a}$ decreases, the effect is more and more pronounced. When it becomes
dominant everywhere outside the core radius of the cluster, the gravitational
potential and therefore the dark matter of the system have little
effect on $\bar{v}_\mathrm{rad}$, 
hence the degeneracy of the profiles in Fig.~\ref{figure3} for $z_{a}$ = 10. Note that as
velocities become radially distributed, light stars need more space to contribute
the same dark mass as a result of the conservation of the phase space volume.
At fixed $M_{2} / M_{1}$ ratio, low mass stars extend further away as the velocity
anisotropy increases. However, there is still a substantial amount of them
inside the visible part of the cluster. When $z_{a}$ = 10 for instance, the dark
mass inside the inner 60 pc is respectively $4.7 \times 10^{4}, 1.3 \times 10^{5},
2.2 \times 10^{5}$ and $2.9 \times 10^{5} \; {\rm M}_{\odot}$ for
a light to heavy stellar mass ratio $M_{2} / M_{1}$ of $1, 3, 6$ and $10$.

The dashed curve that appears in three of the plots of Fig.~\ref{figure3} is the velocity
dispersion profile for a King model with no low mass stars. That case is degenerate
with the entire set of Michie clusters with $z_{a} = 20$, whatever their dark matter
content. The same velocity dispersion profile may therefore be interpreted
either with a one-component isotropic cluster or, alternatively, with an
anisotropic system containing dark matter. If so, the degeneracy may be
lifted by the distribution of surface luminosity. Note however that the
brightness profiles become different far from the center, in a region where
the surface mass density $\Sigma_1$ has already dropped by four orders
of magnitude with respect to its central value. Figures~2 and 3 illustrate actually
the difficulty to estimate the amount of low mass stars potentially concealed
in globular clusters from their velocity dispersion profiles and from the
distributions of their surface luminosities. Both measurements have been
used so far  to determine the mass content of GCs. We therefore conclude that
the same set of observations may be interpreted by quite different models. This
suggests that the actual stellar mass function of many globular clusters needs
to be reinvestigated.

Dark matter in globular clusters should be traced by the three dimensional
velocity distribution of bright stars. We have just shown that the projection
along the line of sight alone does not lead to a unique answer. Another possibility
is the search for tidal tails. If present, dark matter dominates the gravity of
the outskirts of clusters and prevents heavy stars from evaporating
%\cite{Moore96}
(Moore 1996). Tidal tails exist whenever dark matter is not present to play that
inhibition effect. As mentioned by Taillet et al. (1995 and 1996),
%\cite{Taillet95,Taillet96}
the presence of low mass stars should also lead to an infrared halo surrounding the
bright central part. Finally, if the cluster lies against a rich stellar background,
low mass stars should also induce a few gravitational microlensing events.

\begin{acknowledgements}
This work has been carried out under the auspices of the Human
Capital and Mobility Programme of the European Economic Community, under
contract number CHRX-CT93-0120 (DG 12 COMA).
\end{acknowledgements}

\appendix

\section{Useful expansions}

The functions $d(x)$ and $e(x)$ may be expanded as power series of the
variable $x$ when the latter is small. The Dawson's function is given by
\begin{eqnarray}
d(x) &=& x \, - \, A_{1} x^{3} \, + \, A_{1} A_{2} \, x^{5} \, + \,
... \nonumber \\
&+& (-1)^{n} A_{1} A_{2} ... A_{n} \, x^{2n+1} \, + \, ... \;\; ,
\label{DEL_d}
\end{eqnarray}
whereas
\begin{eqnarray}
e(x) &=& x \, + \, A_{1} x^{3} \, + \, A_{1} A_{2} \, x^{5} \, + \,
... \nonumber \\
&+& A_{1} A_{2} ... A_{n} \, x^{2n+1} \, + \, ... \;\; .
\label{DEL_e}
\end{eqnarray}
The constants $A_{n}$ stand for the ratios $2 / (2n+1)$. Whenever
$\alpha \sqrt{u}$ or $\sqrt{u}$ are small compared to 1, expansions
(\ref{DEL_d}) or (\ref{DEL_e}) should be used in expression
(\ref{H_u_a}). If both variables are small at the same time, the expansion
becomes
\begin{eqnarray}
{\cal H} (u,\alpha) & = & \frac{u^{5/2}}{2} A_{1} A_{2}
\left\{ 1 \, + \, A_{3} u \left( 1 - \alpha^{2} \right) \right. \\
&+& A_{3} A_{4} u^{2} \left( 1 - \alpha^{2} + \alpha^{4} \right)
\nonumber \\
& + & A_{3} A_{4} A_{5} u^{3}
\left( 1 - \alpha^{2} + \alpha^{4} - \alpha^{6} \right) \nonumber \\
&+& A_{3} A_{4} A_{5} A_{6} u^{4}
\left( 1 - \alpha^{2} + \alpha^{4} - \alpha^{6} + \alpha^{8} \right)
\nonumber \\
&+& \left. ... \; \right\} \nonumber
\end{eqnarray}
In the same limit, the function ${\cal I}$ may be expanded as
\begin{eqnarray}
{\cal I} (u,\alpha) & = & \frac{u^{7/2}}{2} A_{1} A_{2} A_{3}
\left\{ 1 \, + \, A_{4} u \left( 1 - 2 \alpha^{2} \right) \right.  \nonumber \\
&+& A_{4} A_{5} u^{2} \left( 1 - 2 \alpha^{2} + 3 \alpha^{4} \right)
\nonumber \\
& + & A_{4} A_{5} A_{6} u^{3}
\left( 1 - 2 \alpha^{2} + 3 \alpha^{4} - 4 \alpha^{6} \right) \\
&+& A_{4} A_{5} A_{6} A_{7} u^{4}
\left( 1 - 2 \alpha^{2} + 3 \alpha^{4} - 4 \alpha^{6} + 5 \alpha^{8} \right)
\nonumber \\
&+& \left. ... \; \right\} \nonumber
\end{eqnarray} 
whereas ${\cal J}$ is a sum of three series expansions
\begin{eqnarray}
\left( 1 + \alpha^{2} \right)^{2} \, {\cal J} (u,\alpha) & = &
\frac{u^{9/2}}{15} \alpha^{2} A_{3} A_{4} \;
\left\{ 2 {\cal J}_{a}(u) \right. \nonumber \\
&-& \left( 5 + 7 \alpha^{2} \right) \alpha^{2} {\cal J}_{b}(u \alpha^{2})
\nonumber \\
& + & \left. 9 \left( 1 + \alpha^{2} \right) \alpha^{2}
{\cal J}_{c}(u \alpha^{2}) \right\} \;\; .
\end{eqnarray}
The functions ${\cal J}_{a}$, ${\cal J}_{b}$ and ${\cal J}_{c}$ are respectively
defined by
\begin{eqnarray}
{\cal J}_{a}(x) \; = \; 1 \, + \, A_{5} \, x \, + \, A_{5} A_{6} \,
x^{2} \, +\, A_{5} A_{6} A_{7} \, x^{3} \, + \, ...
\end{eqnarray}
\begin{eqnarray}
{\cal J}_{b}(x) \; = \; 1 \, - \, A_{5} \, x \, + \, A_{5} A_{6} \,
x^{2} \, - \, A_{5} A_{6} A_{7} \, x^{3} \, + \, ...
\end{eqnarray}
while
\begin{eqnarray}
{\cal J}_{c}(x) \; = \; 1 \, - \, A_{4} x {\cal J}_{b}(x) \;\; .
\end{eqnarray}

\end{document}